\RequirePackage[T1]{fontenc}
\documentclass[copyright,creativecommons]{eptcs}

\providecommand{\firstpage}{1}
\providecommand{\publicationstatus}{T. Kutsia and D. Ventura (Ed.): LSFA 2023\\
    Post-proceedings, pp.\ \pageref*{FirstPage}--\pageref{LastPage}}

\usepackage[dvipsnames]{xcolor}
\definecolor{burntumber}{rgb}{21, 96, 189}
\usepackage{stmaryrd,paralist}
\usepackage{xspace,url,hyperref}
\usepackage{cite}
\usepackage{amsfonts,amssymb}
\usepackage{threeparttable,url,breakurl}
\usepackage{listings}
\usepackage[english]{babel} 
\usepackage{tikz}
\usetikzlibrary{cd}

\newtheorem{myteor}{Theorem}[section]

\newtheorem{mylem}{Lemma}[section]

\newcommand{\squarenip}[7][]{
\draw(#2-0.3,#3+0.75) node{#4};
        \draw[-{Stealth[slant=0]}] (#2,#3+1.5) -- (#2,#3);
        \draw(#2+0.75,#3+1.8) node{#5};
        \draw[-{Stealth[slant=0]}] (#2,#3+1.5) -- (#2+1.5,#3+1.5);
        \draw(#2+0.75, #3-0.3) node{#6};
        \draw[-{Stealth[slant=0]}, style = dashed] (#2,#3) -- (#2+1.5,#3);
        \draw(#2+1.8,#3+0.75) node{#7};
        \draw[-{Stealth[slant=0]}, style = dashed] (#2+1.5,#3+1.5) -- (#2+1.5,#3);
        \draw(#2+0.75, #3-0.75) node{#1};
      }

\newcommand{\beluga}{\textsc{Beluga}\xspace}

\newtheorem{@problem}{Exercise}[section]

\newtheorem{@sol}{Solution}[section]

\newtheorem{@axiom}{Axiom}

\lstdefinelanguage{ContextualML}
{
  morekeywords={and, block, case, of, mlam, fn, impossible, let, in, schema,
    some, rec, type, ctype, prop, stratified, inductive, coinductive, LF, if, then,
    else, total},
  keepspaces=true,
  sensitive,
  morecomment=[l]{\%},
  morecomment=[n]{\%\{}{\}\%},
  morestring=[b]"
}[keywords,comments,strings]

\newcommand{\nicettfamilysize}{\footnotesize}
\newcommand{\nicettfamily}{%
    \nicettfamilysize\ttfamily%
}%

\lstdefinelanguage{Beluga}{%
    morecomment = [l][\color{Gray}]{\%},
    morekeywords = [1]{
      LF, type, ctype, schema, rec, proof, as, case, by, of, unbox,
inductive, %
      stratified, some, block, total, mlam, fn, let, in, %
      intros, solve, msplit, suffices, split, by, as, invert, case,
unboxed, %
      impossible, undo, toshow, %
    },%
    keywordstyle = [1]{\bfseries},%
    morekeywords = [2]{
      beta, lm, tla, tbeta, tap, tlm,ap, unit, arr, all, z, s,lam,
      app, v_lam, v_c, s_app_1, abs, tlam, tapp, s_app, s_beta, next,
      halts/m, %
      Unit, Arr, t_lam, t_app, t_c, %
    },%
    keywordstyle = [2]{\bfseries\color{purple!90}},%
    morekeywords = [3]{
      eta_red*, eta_red=, tp, ty, tm, term, nat, val, diff, heigth, not_joinable, step,
      steps, beta_red, eta_red,  mstep, pred,
      mstep_pred, conv, cd, notlam, lam_or_not, halts, Reduce, RedSub,
      ctx, rctx, pctx, pcctx, joinable, strip_prop, conf_prop,
      dia_prop, confl_prop, beta_eta, commute, conf_ord_prop,
      confl_prop_beta, dia_un_prop, conf_un_prop, strip_un_prop,
      conf_beta_eta, oftype, eq, eta*_eta=_joinable, %
      halts_step, bwd_closed, %
    },%
    keywordstyle = [3]{\bfseries\color{RoyalBlue}},%
    alsoletter=/,%
    columns=flexible,%
    sensitive = true,%
    basicstyle=\nicettfamily,%
    mathescape=true%
}%

\newcommand{\EtaSteps}{\mathrel{\,\longrightarrow_\eta\,}}

\begin{document}
\title{More Church-Rosser Proofs in \beluga
}
\author{Alberto Momigliano\thanks{Member of the Gruppo Nazionale Calcolo Scientifico-Istituto Nazionale di Alta Matematica (GNCS-INdAM)} 
 \institute{Dipartimento di Informatica,\\
  Universit\`{a} degli Studi di Milano, Italy}
 \and
Martina Sassella
\institute{Dipartimento di Matematica\\
   Universit\`{a} degli Studi di Milano, Italy}
 }
 \def\authorrunning{A.~Momigliano \& M.~Sassella}
 \def\titlerunning{More Church-Rosser Proofs in \beluga}
\maketitle
\begin{abstract}
 We report on yet another formalization of the Church-Rosser property
 in lambda-calculi, carried out with the proof environment
 \beluga. After the well-known proofs of confluence for
 $\beta$-reduction in the untyped settings, with and without
 Takahashi's complete developments method, we concentrate on $\eta$-reduction
 and obtain the result for $\beta\eta$ modularly. We further extend
 the analysis to typed-calculi, in particular System F. Finally, we
  investigate the idea of pursuing the encoding directly in
 \beluga's meta-logic, as well as the use of \beluga's logic
 programming engine to search for counterexamples.
\end{abstract}

\section{Introduction}\label{sec:intro}

Stop me if you heard this before: the Church-Rosser theorem for
$\beta$-reduction (CR($\beta$)) is a good case study for proof
assistants. 
In fact, Church-Rosser theorems are arguably the most formalized
results in  mechanized meta-theory of deductive systems.

In the beginning, the thrust was to see whether the theorem could be
formalized at all: Shankar's proof in the Boyer-Moore
theorem-prover~\cite{Shankar88} was a break-through and a tour de
(brute) force. A few years later,
Nipkow\cite{DBLP:journals/jar/Nipkow01} made the proof much more
abstract and extended it to $\beta\eta$-reduction. The workhorse
encoding technique was de Bruijn indices and it was a bullet that one just
had to bite.

In the following years, the Church-Rosser theorem became a benchmark
to showcase how to mechanize the variable binding problem. Some
partial data-points, with the understanding that these are not
exhaustive not disjoint: if interested in mirroring the informal
practice of working mathematicians, see the paper by Vestergaard and
Brotherston~\cite{DBLP:conf/rta/VestergaardB01} and the recent work by
Copello et al.~\cite{DBLP:journals/mscs/CopelloST21}.
If you want to reason about $\alpha$-conversion explicitly via
quotients, see Ford and Mason's~\cite{DBLP:journals/tcs/FordM01}. For
the \emph{locally nameless} representation, see McKinna and
Pollack~\cite{DBLP:conf/tlca/McKinnaP93} and its modern
take~\cite{LN}.  Library support is showcased by
\textsc{AutoSubst}~\cite{DBLP:conf/itp/SchaferTS15}.  Nominal
techniques are also well-represented, see the recent proof by Nagele
and al.~\cite{DBLP:journals/corr/NageleOS16}.

We shall use higher-order abstract syntax (HOAS) following up on the
seminal proof of CR($\beta$) by Pfenning in 1992~\cite{Pfenning92cr},
when Twelf was merely Elf. Once the non-trivial issue of
totality-checking was settled, that proof stood as a shining example
of the benefits of HOAS, which we shall not repeat here. Confluence
results did not attract further attention in this community, until
Accattoli, in his 
proof pearl~\cite{DBLP:conf/cpp/Accattoli12}, liberated Huet's Coq
encoding of residual theory~\cite{huet} from its 
 concrete-syntaxed
infrastructure.

\smallskip

So why bother with yet another HOAS-based encoding? 
Our aim was to go beyond $\beta$-reduction and replicate Nipkow's
results in the HOAS setting. Further, to extend it to typed calculi,
with a minimum amount of change.  One way to achieve that is via
\emph{intrinsically-typed terms}~\cite{DBLP:journals/jar/BentonHKM12}
and for the latter, \beluga~\cite{DBLP:conf/cade/PientkaC15} is the
natural (if not only) modern system that natively supports HOAS
together with dependent types.  We were also curious to see if we
could achieve a significant formalization as \beluga's novices, with
just a cursory understanding of its intricate
meta-theory.\footnote{With the partial exception of Kaiser's
  dissertation~\cite{DBLP:conf/rta/KaiserPS17}, all existing \beluga's
  code originates exclusively from Pientka and her students.}
This is part of a more general project of developing a curriculum
to teach the classic theory of the lambda-calculus to
graduate students.

In
passing, we evaluate (in the negative) Accattoli's
suggestion~\cite{DBLP:conf/cpp/Accattoli12} that even in two-level
systems such as \beluga, the specification of the semantics of the
(untyped) lambda-calculus should be carried out directly in the
meta-logic.  
In the Appendix, we explore the concept of \emph{mechanized
  meta-theory model-checking} within the Beluga
framework. This approach aligns with the increasing tendency to
enhance proof assistants with a form of \emph{counterexample}
search. QuickChick serves as one example of this trend~\cite{QC}..


  \smallskip
  
In this short paper, we assume knowledge of confluence in the
lambda-calculus, for which we refer to the concise presentation
in~\cite{selinger2013lecture}. We recall some basic notions; given
binary relations over a set $A$, and their Kleene and reflexive
closure ${(\_)}^*$, ${(\_)}^=$, we define:
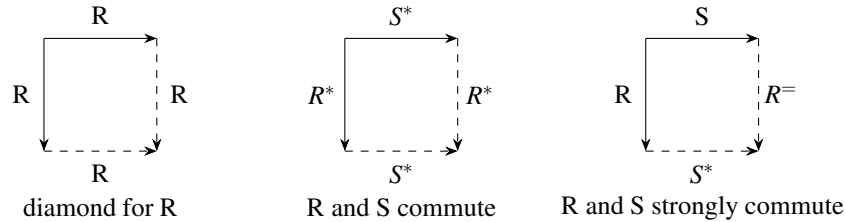
\begin{figure}[h]
  \centering
  \begin{small}
   \begin{center}
     \begin{tikzpicture}
                       \squarenip[diamond for R]{0}{0}{R}{R}{R}{R}
        \squarenip[R and S commute]{4}{0}{$R^*$}{$S^*$}{$S^*$}{$R^*$}
        \squarenip[R and S strongly commute]{8}{0}{R}{S}{$S^*$}{$R^=$}
    \end{tikzpicture}
\end{center}
\end{small}
\end{figure}

\noindent
We say that $R$ is \emph{confluent} if and only it commutes with itself.

We also assume a passing familiarity with \beluga. We simply
recall that the specification of the syntax and judgments of the
system under study is done in (contextual) LF, while theorems are
realized as total functions in \beluga's meta-logic. LF contexts are reified into
first-class objects that can be abstracted and quantified over, being
classified by context \emph{schemas}. First-class substitutions map contexts
to each other realizing properties such as weakening, strengthening
and subsumption. In this development, we have taken care to state
every theorem quantifying over the {smallest} context schema
where it makes sense~\cite{DBLP:journals/mscs/FeltyMP18}.

\section{Sketch of the Formal Development}\label{sec:fomal}

We will only sketch some of the highlights of the encoding, referring the reader to the repository\footnote{\url{https://github.com/martinasassella/More_CR_Proofs_Beluga}}  for most, if not all the details.

\subsection{CR($\beta$)}\label{ssec:beta}

Pientka~\cite{CRporting} ported to \beluga Pfenning's
encoding~\cite{Pfenning92cr} in Twelf of the traditional parallel reduction
proof à la Tait/Martin-Löf.  We did the same for Licata's
proof~\cite{TCD} via Takahashi's \emph{complete
  developments}~\cite{DBLP:journals/iandc/Takahashi95}
This porting was similarly uneventful, save of course for the
improvements that the \beluga brings in w.r.t.\ Twelf. This applies in
particular to the streamlined handling of context reasoning in
contrast to Twelf's rather fragile notion of \emph{regular worlds}. A
detailed comparison of the relative merits of \beluga vs.\ Twelf,
among others, can be found in~\cite{DBLP:journals/jar/FeltyMP15}.

In both proofs, there are \textit{no} technical lemmas about
variables, renamings etc.  We do have to prove that parallel reduction
is stable under substitution, but it is from first principles, meaning
it does nor rely on properties such as weakening and exchange. The
direct proof of the diamond property for parallel reduction has a complex
case-analysis, just as in the informal case.  Takahashi's proof is
based on a \textit{relational} rather than functional encoding of
complete developments, which is fine as the proof only needs totality,
not uniqueness of developments.

For future reference, we list here the HOAS encoding of the syntax of
untyped lambda terms  and of parallel reduction, featuring a crucial
use of hypothetical judgments to internalize the variable cases:
\begin{small}
\begin{lstlisting}
LF term : type =
 | lam : (term -> term) -> term
 | app : term -> term -> term;

LF pred : term -> term -> type =
 | beta : ({x:term} pred x x -> pred (M1 x) (M1' x)) -> pred M2 M2'
                                 -> pred (app (lam M1) M2) (M1' M2')
 | lm : ({x:term} pred x x -> pred (M x) (M' x)) -> pred (lam M) (lam M')
 | ap : pred M1 M1' -> pred M2 M2' -> pred (app M1 M2) (app M1' M2');
\end{lstlisting}
\end{small}

\subsection{CR($\eta$)}\label{ssec:eta}
While CR($\eta$), as well as CR($\beta\eta$)  can be proved via complete
  developments,   here we follow Nipkow's more modular
\emph{commutation} approach. We separately consider $\eta$-reduction
as the congruence over the $\eta$ rule
$\Gamma\vdash \lambda x.M~x \EtaSteps M$, provided
$x\not\in \mathit{FV(M)}$:
$\eta$-reduction (or $\eta$-expansion, if viewed from the right to the left) is encoded in \beluga as the type family
\begin{small}
  \begin{lstlisting}
LF eta_red : term -> term -> type =
 | eta : eta_red (lam \x.(app M x)) M % .. congruence rules as above, omitted
\end{lstlisting}
\end{small}
Note how the
above proviso 
is realized within HOAS as the meta-variable \lstinline!M! \emph{not}
depending on \lstinline!x! in the LF function \lstinline!\x.(app M x)!.
There is nothing ``tricky'' in this encoding, as per Nipkow's discussion (§ 4.3 of
op.cit.). Only one technical lemma is required:
\begin{mylem}(Strengthening) If $\Gamma, x \vdash M \EtaSteps N$, and
  $M$ does not depend on $x$, neither does $N$ and
  $\Gamma \vdash M \EtaSteps N$.
\end{mylem}
Formalizing this takes a little thought, but \beluga's primitives make it relatively easy to state and prove.

The main proof strategy relies on a classic 
result~\cite{DBLP:journals/jacm/Rosen73}
that provides a sufficient condition for \emph{commutation}:

\begin{mylem} (Commutation  -- Hindley-Rosen)\label{commlemma}
  Two strongly commuting reductions commute.
\end{mylem}
Since both LF and \beluga are (roughly) first-order type theories, we
cannot express  operations on abstract relations, in particular
closures, nor can we prove the above lemma once and for all for any
such relation. Hence, we instantiate $R$ and $S$ to $\eta$-reduction:
      
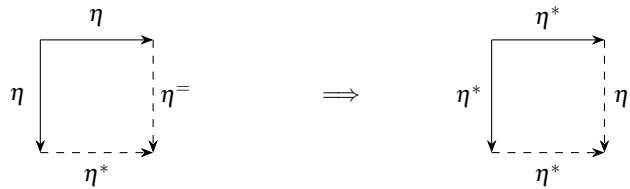
\begin{footnotesize}
    \begin{center}
    \begin{tikzpicture}
    \squarenip[]{0}{0}{$\eta$}{$\eta$}{$\eta^*$}{$\eta^=$} 
    \draw(4,0.75) node{$\Longrightarrow$};
    \squarenip[]{6}{0}{$\eta^*$}{$\eta^*$}{$\eta^*$}{$\eta^*$}
    \end{tikzpicture}
\end{center}
\end{footnotesize}
\vspace{-0.5cm} We define the relevant closures at the LF level, with
families \lstinline|eta_red*| and \lstinline|eta_red=|, and prove that
$\eta$ strongly commutes with itself as the function
\lstinline!square!. Recall that neither LF nor \beluga have
first-class sigma types and therefore existential propositions need to
be encoded separately:
\begin{small}
\begin{lstlisting}
LF eta*_eta=_joinable : term -> term -> type =
 | eta*_eta=_result : eta_red* M1 N -> eta_red= M2 N -> eta*_eta=_joinable M1 M2;

schema ctx = term

rec square : ($\gamma$:ctx) {M : [$\gamma$ |- term]}{M1 : [$\gamma$ |- term]}{M2 : [$\gamma$ |- term]}
          [$\gamma$ |- eta_red M M1] -> [$\gamma$ |- eta_red M M2] -> [$\gamma$ |- eta*_eta=_joinable M1 M2] = ..
\end{lstlisting}
\end{small}
This is proven by induction on \lstinline![$\gamma$ |- eta_red M M1]! and
inversion on \lstinline![$\gamma$ |- eta_red M M2]!, with an appeal to
strengthening when a critical pair is created by the $\eta$ and $\xi$
rules. We then state, in the rather long-winded way that \beluga
requires, the above instance of the Commutation lemma:

\begin{small}
\begin{lstlisting}
LF confl_prop : term -> term -> type =
 | confl_result : eta_red* M1 N -> eta_red* M2 N -> confl_prop M1 M2;

rec commutation_lemma:($\gamma$:ctx)({M: [$\gamma$ |- term]}{M1: [$\gamma$ |- term]}{M2: [$\gamma$ |- term]}
        [$\gamma$ |- eta_red M M1] -> [$\gamma$ |- eta_red M M2] -> [$\gamma$ |- eta*_eta=_joinable M1 M2])
            -> ({M': [$\gamma$ |- term]}{M1': [$\gamma$ |- term]}{M2': [$\gamma$ |- term]}
            [$\gamma$ |- eta_red* M' M1'] -> [$\gamma$ |- eta_red* M' M2'] -> [$\gamma$ |- confl_prop M1' M2']) = ..
\end{lstlisting}
\end{small}
The proof approximates Nipkow's diagrammatic way via a concrete
instance of the \emph{Strip} lemma. In the end, what we decisively
gained in elegance and succinctness with HOAS vs de Bruijn, we
somewhat lose by the limitations of a first-order framework.

\subsection{CR($\beta\eta$)}\label{ssec:be}
We take $\beta\eta$-reduction as the union of $\beta$ and
$\eta$-reductions. Since we have already shown confluence for each
relation separately, it makes  sense to exploit this other classic~\cite{DBLP:journals/jacm/Rosen73}
result:
\begin{mylem}[Commutative Union]\label{communion}
    If $R$ and $S$ are confluent and commute, then $R \cup S$ is confluent.
\end{mylem}

To establish the commutation assumption in the above lemma, we will again appeal
to the Commutation lemma~\ref{commlemma}, hence we start with
a version of the \lstinline|square| function above, with $R:=\beta$ and $S:=\eta$.
The proof requires a strengthening lemma for $\beta$-reduction, as
well as two beautifully clean substitution lemmas for $\eta$ and $\eta^*$. Again, this beauty is short-lived, as we have to replay the diagrammatic proof of
the Commutation lemma with the current instantiation; this is a routine rework of the $\eta$ case, but tedious.

The final ingredient is the proof of lemma~\ref{communion} for
$\beta\eta$, which is entailed by the following steps:
\begin{enumerate}
\item    ${(\beta^* \cup \eta^*)}^* = {(\beta \cup \eta)}^*$;
\item    $\beta^* \cup \eta^*$ satisfies the diamond property;
\item a Strip lemma for the above two relations.
\end{enumerate}

\subsection{Typed Calculi}\label{ssec:sf}

We now switch gears and address confluence  in \emph{typed}
lambda-calculi.  While the main definitions still apply, we must be
mindful to reduce only  well-typed
terms~\cite{selinger2013lecture}. In a dependently typed proof
environment such as \beluga, this can be very elegantly accomplished
using \emph{intrinsically-typed terms}~\cite{DBLP:journals/jar/BentonHKM12},
that is, by ruling out pre-terms and indexing the judgments
under study by well-typed terms only.

To illustrated the idea, we list the specification of   well-typed terms in
the polymorphic lambda-calculus (System F) and a fragment of parallel
reduction (congruence rules for type abstraction and application
together with the two beta rules):
\begin{small}
\begin{lstlisting}
LF ty : type =                    LF tm : ty -> type =
 | arr : ty -> ty -> ty             | lam : (tm A -> tm B) -> tm (arr A B)
 | all : (ty -> ty) -> ty;          | app : tm (arr A B) -> tm A -> tm B
                                    | tlam : ({a:ty} tm (A a)) -> tm (all A)
                                    | tapp : tm (all A) -> {B:ty} tm (A B);

LF pred : tm A -> tm A -> type =
 | tlm : ({a:ty} pred (M a) (M' a)) -> pred (tlam M) (tlam M')
 | tap : pred M M' -> pred (tapp M A) (tapp M' A)
 | beta : ({x:tm A} pred x x -> pred (M1 x) (M1' x)) -> pred M2 M2'
                               -> pred (app (lam M1) M2) (M1' M2')
 | tbeta : ({a:ty} pred (M1 a) (M1' a)) -> pred (tapp (tlam M1) A) (M1' A) ..
\end{lstlisting}
\end{small}
Note how the signature of \lstinline!pred! enforces the invariant that
(parallel) reduction preserves typing.  Since we now have two kinds of
variables, the judgment is hypothetical both on terms and on types. On
the reasoning level, this induces a context schema that accounts for this alternation, namely
\begin{small}
\begin{lstlisting}
schema pctx = some [A:ty] block(x:tm A, v:pred x x) + ty
\end{lstlisting}
\end{small}

What is remarkable is that literally the same proof structure of the
Church-Rosser theorem carries over from the untyped case to the typed
one. To wit, we have formalized Takahashi's style CR($\beta$) for System F and the
proof is \emph{conservative} in a very strong sense: not only do we
use the same sequence of lemmata, but, being \beluga's scripts  explicit
proof-terms, the derivations for the untyped calculus directly \emph{embed}
into the derivations for System F\@: the user just needs to refine the
statements of the theorems with the indexed judgments, on occasion
making some implicit arguments explicit to aid type inference and of
course adding cases for the new constructors.  The proof of CR($\beta$)
for System F consists of around $460$ loc, as opposed to $340$ for the
untyped case.

\subsection{CR($\beta$) at the Meta-Level}
\begin{figure}[t] \label{pred}
\centering
\begin{small}
\begin{lstlisting}
schema rctx = block(x:term, t:pred x x)

rec rpar: {g:rctx}{M: [$\gamma$ |- term]}[$\gamma$ |- pred M M] =
mlam g => mlam M => case [$\gamma$ |- M] of
 | [$\gamma$ |- #p.1] => [$\gamma$ |- #p.2]
 | [$\gamma$ |- lam \x.M'[..,x]] => 
   let [g, b:block(x:term,v:pred x x) |- IH[..,b.1,b.2]] =
     rpar [g, b:block(x:term,v:pred x x)] [g,b |- M'[..,b.1]] in
       [$\gamma$ |- lm \x.\v.IH[..,x,v]]
 | [$\gamma$ |- app M1 M2] => 
   let [$\gamma$ |- IH1] = rpar [g] [$\gamma$ |- M1] in 
   let [$\gamma$ |- IH2] = rpar [g] [$\gamma$ |- M2] in 
       [$\gamma$ |- ap IH1 IH2];

rec rpar: {$\gamma$:ctx}{M : [$\gamma$ |- term]} predM [$\gamma$ |- M] [$\gamma$ |- M] =
mlam g => mlam M => case [$\gamma$ |- M] of
 | [$\gamma$ |- #p] => var (vp _)
 | [$\gamma$ |- lam \x.M'[..,x]] => 
    let h = rpar [g,x:term] [g,x:term |- M'] in
    lm h
 | [$\gamma$ |- app M1 M2] => 
    let h1 = rpar [g] [$\gamma$ |- M1] in
    let h2 = rpar [g] [$\gamma$ |- M2] in
    ap h1 h2;
\end{lstlisting}
\end{small}
\caption{Reflexivity of \lstinline+pred+ vs \lstinline+predM+}
\end{figure}



\label{ssec:ben}
In the two-level approach adopted by \beluga, the syntax and the
semantics of a formal system are specified at the LF level, whereas
reasoning is carried out at the computational level in form of
recursive functions. LF features hypothetical judgments and those give
for free properties of contexts such as exchange, weakening and
substitution; by ``for free'', we mean that they need not to be proved
on a case by case base. Nonetheless, the communication of context
information to the reasoning level requires the definition of possibly
complex context schemas and  relations among them that may complicate
the statements and pollute the proofs. This may be particularly
annoying in case studies such as the untyped lambda calculus, where
contexts do not seem to play a large part, as observed by Accattoli in
his remarkable paper~\cite{DBLP:conf/cpp/Accattoli12} w.r.t.\ the
``cousin'' system Abella.

Since~\cite{DBLP:conf/cade/PientkaC15}, \beluga allows one to define
\texttt{inductive} and \texttt{stratified} relations at the
meta-level. Therefore, we can test Accattoli's proposal w.r.t.\  the
standard proof of CR($\beta$) for the untyped case. This entail keeping
at the LF level only the syntax and move all the other judgments at
the computation level: to wit, parallel reduction has now this
(non-hypothetical) representation:
\begin{lstlisting}
inductive predM : ($\gamma$:ctx) [$\gamma$ |- term] -> [$\gamma$ |- term] -> ctype =
 | var : isVar [$\gamma$ |- M] -> predM [$\gamma$ |- M] [$\gamma$ |- M]
 | betaM : predM [g, x:term |- M1[..,x]] [g, x:term |- M1'[..,x]] -> predM [$\gamma$ |- M2] [$\gamma$ |- M2']
         -> predM [$\gamma$ |- app (lam \x.M1[..,x]) M2] [$\gamma$ |- M1'[..,M2']] % other cases omitted

inductive isVar : ($\gamma$ : ctx) {M: [$\gamma$ |- term]} ctype =
 | vp : {#q: [$\gamma$ |- term]} isVar [$\gamma$ |- #q]
\end{lstlisting}

We have now a separate case for reducing variables, via the
\lstinline!isVar! judgment --- recall that \lstinline|#q| is a
\emph{parameter} variable, ranging over elements of the context. This
is forced upon us by the usual positivity restriction on inductive
types.

The more ominous consequence of this choice is that now establishing
substitution properties for a given judgment requires a proof of
\emph{weakening}, and in turn the latter calls for a proof of context
\emph{exchange}, which is more delicate than expected. First, we must
establish a clear and appropriate definition for determining that the
ordering of a context is irrelevant : binary variable-swapping will
suffice, although we will need to witness the swapping with a
first-class substitution. Unfortunately, showing that reduction is
stable under swapping, which basically amounts to equivariance, is a
hassle: we have to ``explain'' to \beluga's meta-logic what it means
to be a variable. For the gory details, please see directory
\verb+Beta/beta_comp+ in the repository. After that, the main proof
proceeds smoothly in the usual way.


Is the switch to the meta-logic worth it in \beluga? Not really. The
elegance of Accattoli's approach stems from meta-level contexts in
Abella being equivariant, and this preempts any issue with
exchange. On the other hand, the idea works only for contexts that
track ``bare'' bound variables, making it suitable just for (certain)
untyped calculi.  While \beluga can easily overcome this via
intrinsically typed terms, it seems that we have to rebuild a sort of
de Bruijn infrastructure specific to each case study; furthermore, we
had to struggle to prove that a substitutive judgment promoted to the
meta-level is preserved by swapping. This seems too steep of a price
for the mostly cosmetic improvements shown in Figure 1, which
depicts the proof of reflexivity of parallel reduction in the two
flavors. 

\section{Conclusions}\label{sec:conc}

Beluga's support for HOAS, paired with sophisticated context
reasoning, makes the development of the traditional proof of
CR($\beta$) very elegant and devoid of technical lemmas foreign to the
mathematics of the problem. Since specifications can be
dependently-typed, the extension of the proof from the untyped to the
typed case is conservative. It would be easy to encode other proof techniques such
as establishing confluence for the simply typed case using Newman's lemma and
existent  SN proofs in \beluga~\cite{DBLP:journals/jfp/AbelAHPMSS19} or
other classical results such as $\eta$-postponement, standardization, and residuals
theory as in~\cite{DBLP:conf/cpp/Accattoli12}.

On the flip side, \beluga (and LF) not
allowing quantification over relations prevents us from a more
succinct development via abstract rewriting system as per Nipkow's
account, see the repetitions around the Commutation lemma. Combining
natively HOAS and a full Agda-like type theory is under active
research~\cite{DBLP:conf/lics/PientkaT00Z19}. Reader, you may expect
more Church-Rosser proofs in the future.

\paragraph{Acknowledgments}
Thanks to Brigitte Pientka for her help with the development.
\bibliographystyle{eptcs}
\bibliography{b}
\appendix
\section{Appendix: Counterexamples Search}\label{sec:ces}

The theory of confluence is rife with counterexamples.  
It is a well-known fact that (single step) $\beta$-reduction does
\emph{not} satisfy the diamond property, since redexes can be
discarded or duplicated, and this is why the notion of parallel
reduction is so useful. It would be nice, both for research and
educational purposes, for a proof environment to assist us in
\emph{refuting} unprovable conjectures and witnessing such
counter-examples.
A lightweight approach to this endeavor is \emph{property-based
  testing} (PBT), see for example the integration of \emph{QuickChick}
within Coq~\cite{QC}.  If we view a property as a logical formula
$\forall x : \tau. P(x) \supset Q(x)$, providing a counter-example
consists of negating the property, and searching for a proof of
$ \exists x : \tau. P(x) \land \neg Q(x) $.  This points to a logic
programming solution, where the specification is a fixed set of
assumptions and the negated property is the goal. A full
proof-theoretic reconstruction of PBT has been presented
in~\cite{Blanco19} and can be adapted to \beluga, which in the Twelf's tradition
 has a logic programming engine built-in.
 
 To exemplify, let us search for a counter-example to
 diamond($\beta$). First, we need to state the conjecture that we want
 to test, namely the negation of $M_1 \leftarrow M\rightarrow M_2$
 entails $\exists N,\ M_1 \rightarrow N\leftarrow M_2$. This needs a
 little care, since \beluga does not have negation --- the usual
 solution in logic programming, i.e.\ \emph{negation-as-failure}, is
 incompatible with \beluga's foundations: indeed, what could be the
 proof term witnessing a proof failure? One
 solution~\cite{Momigliano00} is to state in the positive when two
 terms are \emph{non}-joinable: i.e., when they are different and if
 they reduce in one step, they do not reduce to the same term. This
 requires a notion of \emph{inequality} for lambda terms (simplified
 here from the actual code):
\begin{small}
\begin{lstlisting}
LF diff : term -> term -> type =                       
 | dal : diff (lam _) (app _ _)
 | dla : diff (app _ _) (lam _)
 | da1 : diff E1 F1 -> diff (app E1 E2) (app F1 F2)
 | da2 : diff E2 F2 -> diff (app E1 E2) (app F1 F2)
 | dll : ({x:term} -> diff (M x) (N x)) -> diff (lam M) (lam N);

LF not_joinable : term -> term -> type =
| nj : diff M1 M2 -> step M1 P1 -> step M2 P2 -> diff P1 P2 -> not_joinable M1 M2;
\end{lstlisting}
\end{small}

The second ingredient is a \emph{generator} for terms. For the sake of
this paper we do not implement the full architecture
of~\cite{Blanco19}, but simply program an exhaustive height-bounded
term generator. We also show the test harness predicate, combining
generation with the to-be-tested conjecture:
\begin{small}
\begin{lstlisting} 
LF heigth : nat -> term -> type =
 | h1 : heigth H M -> heigth H N -> heigth (s H) (app M N)
 | h2 : ({x:term} ({h:nat} heigth h x) -> heigth H (M x)) -> heigth (s H) (lam M);

LF gencex : nat -> term -> term -> term -> type =
| cx : not_joinable M1 M2 -> step M M1 -> step M M2 -> hei I M -> gencex I M M1 M2;
\end{lstlisting}
\end{small}
A query to \beluga's logic programming engine with a bound of $3$ will
generate the following (pretty-printed) counterexample to diamond($\beta$), namely 
$\mathtt{M = }(\lambda x.\ x\ x) (I\ I)$, for $I$ the identity combinator. In fact, it
steps to $ (I\ I) (I\ I)$ and $(\lambda x.\ x\ x) \ I$.

Another application is witnessing the failure of diamond($\eta$) in a
typed calculus with unit and surjective pairing. Here we exploit
intrinsically-typed terms to encode typed $\eta$-reduction and obtain
well-typed term generators for free. All the details in the repository, under
directory \texttt{PBT}.


\end{document}